\documentclass{article}
\usepackage{hyperref}
\usepackage{bm}

\usepackage{rotating}
\usepackage{booktabs}

\usepackage{amssymb}	% Extra maths symbols
\usepackage{authblk}
\usepackage{graphicx}
\usepackage{booktabs}
\usepackage{tabularx, booktabs, makecell, caption}
\usepackage{siunitx}
\usepackage{float}\usepackage[
singlelinecheck=false % <-- important
]{caption}

\begin{document}
\title{\bf Kerr-like black holes shadow surrounded by dark matter halos: Comparison between various dark matter profiles}
\author{{Malihe Heydari-Fard%
\thanks{Electronic address: \href{mailto:heydarifard@qom.ac.ir}{heydarifard@qom.ac.ir}} and Mohaddese Heydari-Fard\thanks{Electronic address: \href{mailto:mo.heydarifard@qom.ac.ir}{mo.heydarifard@qom.ac.ir}} }\\ {\small \emph{ Department of Physics, The University of Qom, 3716146611, Qom, Iran}}
}

\maketitle
\begin{abstract}
The study of shadows of static and rotating black holes immersed in various dark matter halo profiles has gained significant attention in recent years. In this paper, we consider the static black hole solution immersed in three dark matter halo profiles —King, Hernquist, and Moore— and, by using the Newman–Janis algorithm, obtain the corresponding rotating black hole solutions. The main goal of the paper is to study the influence of the characteristic density, characteristic radius, and the spin parameter on the inner and outer horizons, the ergoregion, and the shadow radius, and to compare the results with Kerr black holes and other dark matter profiles. Our findings indicate that the shadow radius of rotating black holes immersed in dark matter increases compared to that of the Kerr black hole in the absence of dark matter; however, the impact on the size and shape of the shadow is negligible for King, Hernquist, and Moore profiles, and for all other dark matter halo profiles considered, and is independent of whether the halo is cuspy or cored. Therefore, at least for these Kerr-like black holes, the black hole shadow is not a suitable tool for distinguishing the nature of the dark matter distribution in galactic centers.
\vspace{5mm}
\end{abstract}

\section{Introduction}
\label{1-intro}

The problem of dark matter (DM) and dark energy are important topics in cosmology and observational astrophysics that have attracted considerable attention in recent years and have been the subject of extensive studies. More and more observational data including the rotation curves of spiral galaxies \cite{Rubin1970, Corbelli:1999af}, the globular clusters \cite{Clowe:2006eq}, formation of large-scale structure \cite{Davis:1985rj}, the cosmic microwave background radiation, baryon acoustic oscillations \cite{WMAP:2010qai, Planck:2013pxb}, and gravitational lensing \cite{Massey:2010hh, Vegetti:2023mgp} indicate the existence of DM in our universe. Based on these observational data, astronomers have proposed a variety of DM models to study DM. In particular, observations of cosmic microwave background radiation suggest that our universe is composed of \%26.8 DM and \%68.3 of dark energy \cite{Planck:2013pxb}. The DM typically forms halo structures in galaxies and galaxy clusters, and influences their behavior. Furthermore, the DM widely influences the motion of stars and particles in galaxies and galaxy clusters by creating a strong gravitational potential \cite{26}-\cite{28}.

On the other hand, in recent years, specifically since the detection of gravitational waves from binary black hole (BH) merger \cite{LIGO:2017dbh}, the study of BHs has been at the forefront of researcher's attention. From an observational perspective, the BHs play a fundamental and crucial role in studying high-energy astrophysical phenomena, such as accretion disks, relativistic jets, and gamma-ray bursts. In addition to stellar-mass BHs, which help us to understand the final stages of stellar evolution and the dynamics of compact objects, astronomers have found observational evidence for existence of supermassive BHs at the centers of galaxies \cite{Ghez:1998ph}, which play an important role in the formation and evolution of galaxies. For example, by observing the S2 star orbiting the center of the Milky Way galaxy, they found that the compact object at the center of galaxy is a supermassive BH with a mass of order $4.3\times10^{6}M_{\odot}$ \cite{Schodel:2002py}. Moreover, the event horizon telescope (EHT) released in April 2019, the first image of supermassive BH at the center of the elliptical galaxy Messier 87 \cite{EventHorizonTelescope:2019dse}, and later in 2022 the image of Sgr A* at the center of our own Milky Way \cite{EventHorizonTelescope:2022wkp}. Therefore, today we know that BHs exist in our universe. Synge \cite{Synge} was the first to study the BH shadow for the Schwarzschild space-time and then Luminet generalized his work to the case of a Schwarzschild BH surrounded by thin accretion disk \cite{Luminet}. The shadow of rotating Kerr BH and Reissner-Nordstrom BH was studied in Refs.\cite{Bardeen} and \cite{Kumaran:2022soh}, respectively. Also, some authors have attempted to test gravity theories using observations of the shadow of M87 and Sgr A* BHs.

Given the existence of supermassive BHs at the centers of galaxies, and the fact that the majority of matter in galaxies is DM, extensive studies have recently been done on BHs immersed in dark halos to explore the possible effects of DM on the properties of BHs in galactic centers: such as null geodesics and BH shadow \cite{Konoplya:2019sns}-\cite{Faraji:2025efg}, accretion disks and time-like geodesics \cite{Heydari-Fard:2022xhr}, binary BH merger processes \cite{Bamber:2022pbs, Karydas:2024fcn}, chaos and BH thermodynamics \cite{Cao:2021dcq}-\cite{Xu:2016ylr}, quasi-normal modes \cite{Jusufi:2019ltj}-\cite{Liu:2024xcd}, and BH echoes \cite{Liu:2021xfb}. In this context, we investigate the shadow of a rotating BHs surrounded by DM halos.

Although numerical simulations have been used to analyze the growth of BHs within DM halos, obtaining an analytical solution for BHs surrounded by dark halos remains challenging. This difficulty arises from the unknown behavior of DM particles when interacting with BHs. To obtain an exact BH solution within DM halo, there are two different approaches: In the first approach, using the density profile of dark halo, the space-time geometry describing a BH within a galactic halo can be obtained. Indeed, the authors generalize the “Einstein cluster” to include a central BH. Initially, they consider a spherical distribution of particles orbiting the center and then add a central mass to this system. In this manner, the hydrostatic equilibrium equations for the BH-halo system are simultaneously and self-consistently solved. This method naturally leads to an energy-momentum tensor with tangential pressure, and vanishing radial pressure. Finally, by employing the asymptotically flatness condition and solving the Einstein field equations, the mass function can be obtain which in small scales reduce to the BH mass and in the large scales to the galaxy mass. This approach was first introduced by Cardoso et al. by considering Hernquist-type density distribution for DM halo \cite{Cardoso}. In second approach, to get the space-time metric of BH immersed in DM halo, the DM density profile $\rho(r)$ and the central BH mass $M$ can be combined to the energy-momentum tensor in the Einstein field equation. It is shown that the effective space-time metric can be decomposed into two parts $f(r)=f_{\rm DM}(r)-\frac{2M}{r}$, in which $f_{\rm DM}(r)$ describes the DM halo and $-\frac{2M}{r}$ is the effects of supermassive BH at the center of galaxy \cite{Xu2018}. Using these approaches, the metric of supermassive BHs surrounded by different DM halos was obtained in many references \cite{Xu2018}-\cite{Qiao:2024ehj} and is summarized in Table 1.

Among various DM halo profiles presented in Table 1, only the King, Hernquist and Moore DM halos have not been investigated so far. In this work,
we consider the static spherically symmetric BH immersed in a King \cite{Al-Badawi:2026cjx}, Hernquist and Moore profiles and then using Newman-Janis algorithm we extend this solution to the case of rotation, and study the effects of DM halo parameters on event horizon. ergoregion and the BH shadow radius. Finally, we compare the effect of DM on the shadow of different BHs. Our results indicate that for all types of DM profiles---cuspy or cored---the effect of DM on the shadow is very small. The structure of the paper is as follows. In Sec.~\ref{2-BH} , we construct a solution for a rotating BH surrounded by a King, Hernquist and Moore DM halos using Newman-Janis algorithm. In addition, we study some physical properties for such rotating BHs including the event horizon of BH and the shape of ergoregion. The null geodesics, circular orbits and the BH shadow are studied in detail in Sec.~\ref{3-null} and Sec.~\ref{4-shadow}. Finally, we end up with our conclusions in Sec.~\ref{5-results}.

%&&&&&&&&&&&&&&&&&&&&&&&&&&&&&&&&&&&&&&&&&&&&&&&&&&&&&&&&&&&&&&&&&&&&&&&&&&&&&&&&&&&&&&&&&&&&&&&&&&&&&&&&&&&&&&&&&&&&&&&&&&&&&&&&&&&&&&&&&&&&&&&&&&&&&&&&&&&&&&&&

\section{Kerr-like BHs in DM halos}
\label{2-BH}

In this section, we give the space-time metric for rotating BHs surrounded by DM halos. The DM distributions at the center of galaxies are usually described by several astrophysical models, such as King, Burkert, Hernquist, NFW, and Moore models. The general DM density profile is given by \cite{Dekel-Zhao}
\begin{equation}
\rho(r) = \frac{\rho_s}{\left(\frac{r}{r_s}\right)^{\gamma}\left[1+\left(\frac{r}{r_s}\right)^{\alpha}\right]^{\frac{\beta-\gamma}{\alpha}}},
\label{1}
\end{equation}
where is known as the Dekel-Zhao profile, and \( \rho_s \) and \( r_s \) are the characteristic density and characteristic radius, respectively. Moreover, \( \alpha \), \( \beta \), and \( \gamma \) parameters are variables used to refer to a specific model. In Table 1, we have listed some different density profiles that are constructed by choosing different values of \( \alpha \), \( \beta \), and \( \gamma \) in the density profile (\ref{1}). Among various DM halo profiles in Table 1, the Kerr-like BH solutions of the King, Hernquist and Moore profiles and their shadows has not been studied yet; so in the present work we will consider these profiles. The DM distributions of the King ($\alpha=2$, $\beta=3$, $\gamma=0$), Hernquist ($\alpha=1$, $\beta=4$, $\gamma=1$), and Moore ($\alpha=3/2$, $\beta=3$, $\gamma=3/2$) models are given by \cite{King}, \cite{Hernquist}, \cite{Moore:1999gc}

\begin{equation}
\rho_{\rm King}(r)=\frac{\rho_{s}}{\left[1+\left(\frac{r}{r_s}\right)^2\right]^{3/2}},
\label{2}
\end{equation}

\begin{equation}
\rho_{\rm Hernquist}(r)=\frac{\rho_{s}}{\left(\frac{r}{r_s}\right)\left[1+\left(\frac{r}{r_s}\right)\right]^{3}},
\label{n3}
\end{equation}

\begin{equation}
\rho_{\rm Moore}(r)=\frac{\rho_{s}}{\left(\frac{r}{r_s}\right)^{3/2}\left[1+\left(\frac{r}{r_s}\right)^{3/2}\right]}.
\label{n4}
\end{equation}
The static spherically symmetric BH solutions surrounded by the King, Hernquist and Moore DM halos are given by \cite{Al-Badawi:2026cjx}, \cite{Jumaniyozov:2025xxh}, \cite{Liu:2023xtb}
\begin{equation}
ds^2=-f(r)dt^2+\frac{dr^2}{f(r)}+r^2\left(d\theta^2+\sin^2\theta d\varphi^2\right),
\label{3}
\end{equation}
where the redshift functions can be expressed as
\begin{equation}
f_{\rm King}(r)=\left(\sqrt{1+\frac{r^2}{r_s^2}}+\frac{r^2}{r_s}\right)^{-\frac{8\pi\rho_{s}r_{s}^3}{r}}-\frac{2M}{r},
\label{4}
\end{equation}

\begin{equation}
f_{\rm Hernquist}(r)= \left( r_s^2 + r^2 \right)^{4\pi\rho_s r_s^2} \exp\left[ \frac{8\pi\rho_s r_s^3 \arctan\left(\frac{r}{r_s}\right)}{r}\right]-\frac{2M}{r},
\label{n7}
\end{equation}

\begin{equation}
f_{\rm Moore}(r)=\exp\left[{\frac{16\pi \rho_{s}r_s^2}{\sqrt{3}} \arctan\left(\frac{2\sqrt{{r}/{r_s}}-1}{\sqrt{3}}\right)}\right] \left(1+(\frac{r}{r_s})^{3/2}\right)^{-\frac{16\pi\rho_{s}r_s^3}{3r}} \left(\frac{1+\frac{r}{r_s}-\sqrt{\frac{r}{r_s}}}{1+\frac{r}{r_s}+2\sqrt{\frac{r}{r_s}}}\right)^{-\frac{8\pi\rho_{s}r_s^2}{3}}- \frac{2M}{r},
\label{n8}
\end{equation}
where $M$ is the BH mass. Now, using the revised Newman-Janis method, we obtain the Kerr-like BH solutions surrounded by these DM halos which in the Boyer-Lindqist coordinates are given by
\begin{eqnarray}
ds^2&=&-\left(\frac{\Delta-a^2\sin^2\theta}{\Sigma}\right)dt^2+\frac{\Sigma}{\Delta}{dr^2}+\Sigma d\theta^2\nonumber\\
&+&\frac{2a\sin^2\theta}{\Sigma}\left(\Delta-(r^2+a^2)\right)dt d\varphi\nonumber\\
&+&\frac{\sin^2\theta}{\Sigma}\left[(r^2+a^2)^2-\Delta a^2\sin^2\theta\right]d\varphi^2,
\label{5}
\end{eqnarray}
with
\begin{equation}
\Delta(r) = r^2+a^2-2 m(r) r, \quad \Sigma(r,\theta) = r^2+a^2\cos^2\theta,
\label{6}
\end{equation}
and
\begin{equation}
m_{\rm King}(r)=M+\frac{r}{2}\left[1-\left({\frac{r}{r_s}}+\sqrt{1+\left(\frac{r}{r_s}\right)^2}\right)^{\frac{-8\pi\rho_s r_s^3}{r}}\right],
\label{7}
\end{equation}

\begin{equation}
m_{\rm Hernquist}(r)=M+\frac{r}{2}\left[1-\left( r_s^2 + r^2 \right)^{4\pi\rho_s r_s^2}\exp\left[\frac{8\pi\rho_s r_s^3 \arctan\left(\frac{r}{r_s}\right)}{r} \right]\right]
\label{7}
\end{equation}

\begin{equation}
m_{\rm Moore}(r)=M+\frac{r}{2}\left[1-\exp\left[{\frac{16\pi \rho_{s}r_s^2}{\sqrt{3}} \arctan\left(\frac{2\sqrt{{r}/{r_s}}-1}{\sqrt{3}}\right)}\right] \left(1+(\frac{r}{r_s})^{3/2}\right)^{-\frac{16\pi\rho_{s}r_s^3}{3r}} \left(\frac{1+\frac{r}{r_s}-\sqrt{\frac{r}{r_s}}}{1+\frac{r}{r_s}+2\sqrt{\frac{r}{r_s}}}\right)^{-\frac{8\pi\rho_{s}r_s^2}{3}}\right].
\label{7}
\end{equation}
Clearly, in the absence of DM, i.e. $\rho_s=0$, the above metrics reduce to the Kerr metric and also for $a=0$ they reduce to the static line elements in Eqs.~\ref{4} -\ref{n8}.

\begin{figure}[H]
\centering
\includegraphics[width=3.0in]{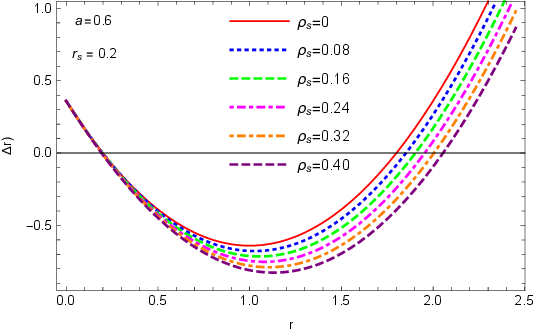}
\includegraphics[width=3.0in]{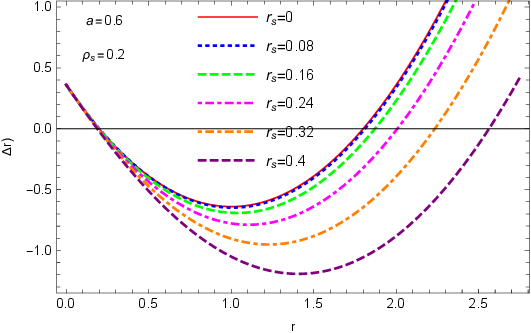}\\
\includegraphics[width=3.0in]{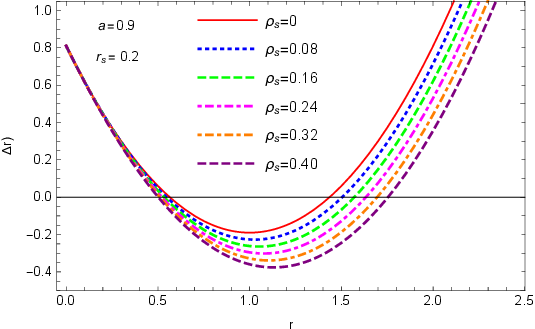}
\includegraphics[width=3.0in]{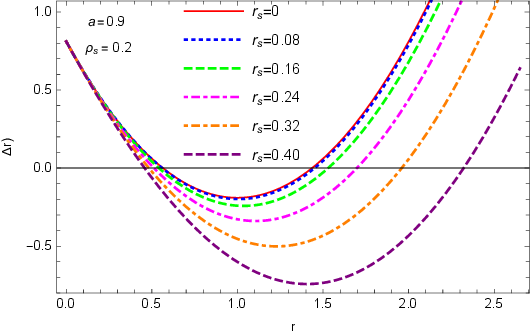}
\caption{\footnotesize The behavior of function $\Delta(r)$ for different values of DM parameters $\rho_s$ and $r_s$, with $M=1$. The rotation parameter is set to $a=0.6$ and $a=0.9$ in the top and bottom rows, respectively. In each panel the solid red curve corresponds to the Kerr BH in the absence of DM.}
\label{1-delta}
\end{figure}

\begin{figure}[H]
\centering
\includegraphics[width=3.0in]{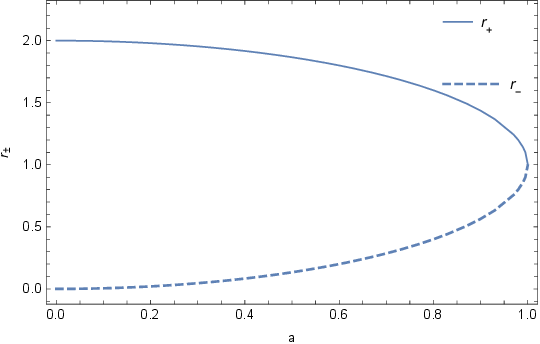}
\caption{\footnotesize The behavior of the inner ($r_-$) and outer ($r_+$) horizons as a function of spin parameter $a$ for $\rho_s=0.04$, $r_s=0.02$ and $M=1$.}
\label{2-horizon}
\end{figure}

Here, it should be noted that since the physical results for the Hernquist and Moore models are similar to that of the King DM halo model, in what follows, for simplicity, we will focus our attention on the King profile. From the condition $g^{rr}=\Delta=0$ one can find the location of horizons. The analysis of the zeros of $\Delta=0$ shows that for some values of the parameters $\rho_s$ and $r_s$, there are two horizons, namely the inner Cauchy horizon, $r_-$, and the outer event horizon, $r_+$, so that $r_-<r_+$. However, for a fixed value of the spin parameter $a$, there are a critical value of DM parameters, for which the two horizons coincide, $r_-=r_+$, corresponding to an extremal BH. We have plotted the behavior of the $\Delta((r)$ function for different values of DM halo parameters $\rho_s$ and $r_s$, with $a=0.6$ and $a=0.9$ in Fig.~\ref{1-delta}. It is easy to see that for a fixed value of spin parameter $a$, increasing both BH parameters $\rho_s$ and $r_s$ causes an increase of the event horizon radius so that Kerr BH with $\rho_s=0$ has the smallest event horizon. Moreover, comparing top and bottom panels of figure shows that by increasing $a$ the BH event horizon radius shifts to lower radii. The dependence of the event horizon on the spin parameter for a given value of DM halo parameters is also shown in Fig.~\ref{2-horizon}. As was mentioned before, by increasing $a$, the outer horizon decreases while the inner one increases and then for an extremal value of $a$ two horizons coincide.

\begin{figure}[H]
\centering
\includegraphics[width=3.0in]{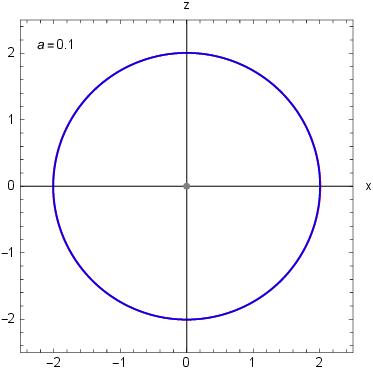}
\includegraphics[width=3.0in]{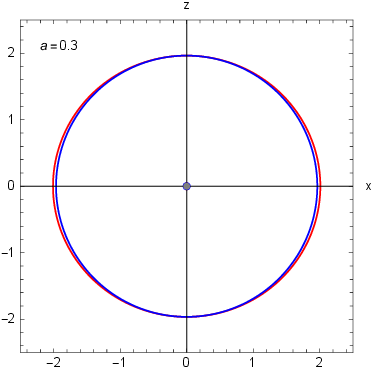}\\
\includegraphics[width=3.0in]{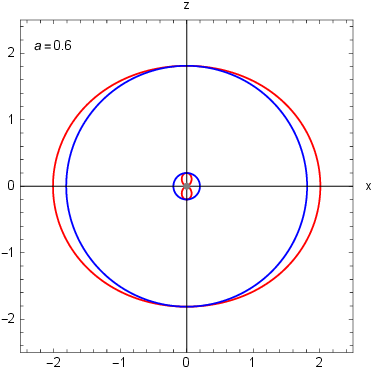}
\includegraphics[width=3.0in]{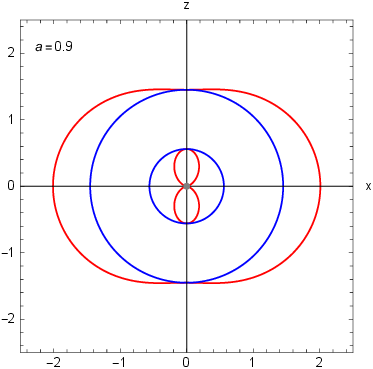}
\caption{\footnotesize The shape of ergosphere of the Kerr-like BH in a King DM halo with different rotation parameter $a$ in the $xz$-plane. The DM halo parameters is set to $\rho_s=0.2$ and $r_s=0.08$, with $M=1$.}
\label{3-ergo}
\end{figure}

Next, we study the shape of the ergosphere of the Kerr-like BH in a King DM halo. As we know,  the ergosphere is a region between the event horizon and the static limit; i.e.,
\begin{equation}
r_+<r<r_{\rm sl},
\label{8}
\end{equation}
where $r_{\rm sl}$ is the radius of the static limit which in the Boyer–Lindquist coordinates is given by condition $g_{tt}=0$,
\begin{equation}
r^2\left({\frac{r}{r_s}}+\sqrt{1+\left(\frac{r}{r_s}\right)^2}\right)^{\frac{-8\pi\rho_s r_s^3}{r}}-2Mr+a^2=0.
\label{9}
\end{equation}
From Fig.~\ref{3-ergo} we see that for a given $\rho_s$ and $r_s$ by increasing spin parameter $a$, the outer horizon and outer infinite redshift surface decrease, while the inner horizon and inner infinite redshift surface increase.

%&&&&&&&&&&&&&&&&&&&&&&&&&&&&&&&&&&&&&&&&&&&&&&&&&&&

%&&&&&&&&&&&&&&&&&&&&&&&&&&&&&&&&&&&&&&&&&&&&&&&&&&&&&&&&&&

\section{Null geodesics in galactic DM halo}
\label{3-null}

In this section, we are interested to study the photon motion around Kerr-like BHs in a King DM halo, which can be expressed with Hamilton-Jacobi equations
\begin{equation}
\frac{\partial S}{\partial\lambda}= -\frac{1}{2} g^{\mu\nu}\frac{\partial S}{\partial x^{\mu}}\frac{\partial S}{\partial x^{\nu}},\label{10}
\end{equation}
where $S$ is the Jacobi action, $\lambda$ is an affine parameter and $g^{\mu\nu}$ is the inverse metric of space-time. Then, one can separate the Jacobi function in the following form
\begin{equation}
S = -Et + L\varphi + S_r(r) + S_\theta(\theta), \label{11}
\end{equation}
where $E$ and $L$ are the conserved quantities correspond to the energy and angular momentum of the photons, respectively. Also, functions $Sr(r)$ and $S_\theta(\theta)$ are arbitrary functions which depend only on the specified coordinates. Taking (\ref{5}) and (\ref{11}) into account, Eq.~\ref{10} can be divided into following two equations
\begin{equation}
\Delta^2\left(\frac{dS_r}{dr}\right)^2 = \left[ (r^2 + a^2)E - aL \right]^2 - \Delta \left[ {\cal K} + (L - aE)^2  \right], \label{12}
\end{equation}
\begin{equation}
\left(\frac{dS_\theta}{d\theta}\right)^2 = {\cal K} - L^2 \cot^2\theta + a^2 E^2 \cos^2\theta, \label{13}
\end{equation}
where ${\cal K}$ is the Carter constant. Now, from $p_{\nu}=\frac{\partial S}{\partial x^{\nu}}$, we can obtain the geodetic equations of photons as
\begin{eqnarray}
\Sigma \dot{t} &=& \frac{E}{\Delta} \left[ (r^2 + a^2)(r^2 + a^2-a\xi) -\Delta a (a \sin^2 \theta-\xi) \right],\label{14}\\
\Sigma \dot{\varphi} &=& \frac{E}{\Delta} \left[ a(r^2 + a^2+a\xi \csc^2 \theta)-\Delta (a+\xi\csc^2 \theta)\right],\label{15}\\
\Sigma \dot{r} &=& \pm\sqrt{ {\cal R}(r)},\label{16}\\
\Sigma\dot{\theta}&=&\pm\sqrt{ \Theta(\theta)},\label{17}
\end{eqnarray}
with
\begin{eqnarray}
{\cal R}(r) &=& E^2\left[ (r^2 + a^2-a \xi)^2 - \Delta [(\xi-a)^2  +\eta]\right],\label{18}\\
\Theta(\theta) &=& E^2\left[\eta-\xi^2 \cot^2\theta +a^2\cos^2\theta\right],\label{19}
\end{eqnarray}
where we have defined new variables $\eta=\frac{{\cal K}}{E^2}$ and $\xi=\frac{L}{E}$, and dot in Eqs.~\ref{14}-\ref{17} denotes the derivative with respect to the affine parameter. Next, we focus on the motion of photons in circular orbits. In order to, we analyze the presence of unstable circular orbits around the BH by rewriting the radial geodesic Eq.~\ref{14} in terms of the effective potential, $V_{\rm eff}$, as
\begin{equation}
\Sigma ^2\dot{r}^2+ V_{\rm eff}(r)=0,\label{20}
\end{equation}
where the effective potential is
\begin{equation}
\frac{V_{\rm eff}(r)}{E^2} = -\left(r^2 + a^2-a \xi \right)^2 + \Delta \left[(\xi-a)^2  +\eta \right].\label{21}
\end{equation}
The conditions for unstable circular orbits are given by
\begin{equation}
V_{\rm eff}(r)|_{r=r_{\rm ph}} = 0, \quad V_{\rm eff}'(r)|_{r=r_{\rm ph}} = 0,\label{22}
\end{equation}
where $r_{\rm ph}$ represents the radius of the unstable photon orbits, and by solving these equations, we find $\eta$ and $\xi $ as
\begin{eqnarray}
\eta &=& \frac{r^2 \left[16 a^2 \Delta(r) - 16 \Delta(r)^2 + 8 r \Delta(r) \Delta'(r) - r^2 \Delta'(r)^2\right]}{a^2 \Delta'(r)^2}|_{r=r_{\rm ph}},\label{23}\\
\xi &=& \frac{- 4 r \Delta(r) + (r^2 + a^2)\Delta'(r)}{a \Delta'(r)}|_{r=r_{\rm ph}},\label{24}
\end{eqnarray}
where the prime shows the derivative with respect to the radial coordinate $r$. Clearly, in the limiting case $\rho_s=0$ we recover the results for the Kerr BH
\begin{eqnarray}
\eta &=& \frac{r^3 \left[4 M a^2-r (r- 3M)^2 \right]}{a^2 (M-r)^2},\label{25}\\
\xi &=& \frac{r^2 (r-3M)+a^2 (r+M)}{a(M-r)}.\label{26}
\end{eqnarray}

\section{Shadow of Kerr-like BHs in the DM halos}
\label{4-shadow}

\begin{figure}[H]
\centering
\includegraphics[width=2.80in]{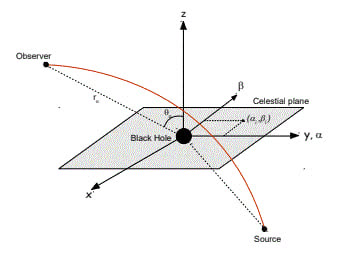}
\caption{\footnotesize Celestial coordinates \cite{Vazquez:2003zm}.}
\label{alpha}
\end{figure}

Now we proceed to find the BH shadow of the Kerr-like BHs in the King DM halo and compare it with that of the Kerr BH in the absence of DM. To this end, we consider an observer at the position $(r_{\rm o},\theta_{\rm o})$ (see Fig.~\ref{alpha}), where $r_{\rm o}\rightarrow\infty$ and $\theta_{\rm o}$ is the inclination angle of the observer. Then, the celestial coordinates read as
\begin{eqnarray}
\alpha &=& \lim_{r_{\rm o} \to \infty} \left( -r_{\rm o}^2 \sin \theta_{\rm o} \frac{d\varphi}{dr} \right),\label{27}\\
\beta &=& \lim_{r_{\rm o} \to \infty} \left(r_{\rm o}^2 \frac{d\theta}{dr}\right).\label{28}
\end{eqnarray}

\begin{figure}[H]
\centering
\includegraphics[width=2.95in]{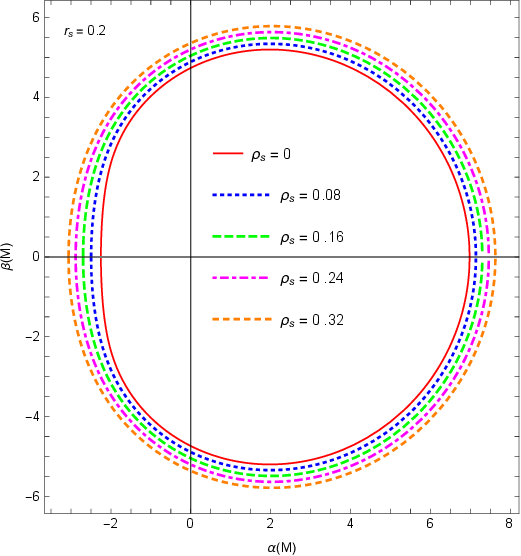}
\includegraphics[width=3.0in]{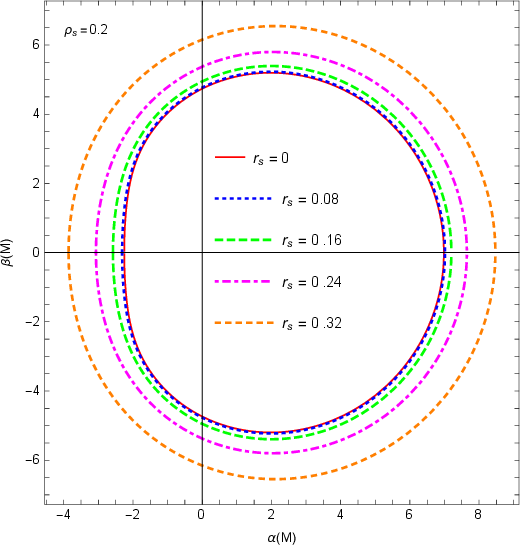}
\caption{\footnotesize The shape of shadow of the Kerr-like BH in a King DM halo for different DM densities $\rho_s$ with $r_s=0.2$ (left panel), and for different halo radius $r_s$ with $\rho_s=0.2$ (right panel). We set $a=0.99$, $\theta_{\rm o}=\frac{\pi}{2}$ and $M=1$. In each panel the solid red curve corresponds to the Kerr BH in the absence of DM.}
\label{4-shadow1}
\end{figure}

Substituting $\frac{d\theta}{dr}$ and $\frac{d\varphi}{dr}$ from Eqs.~\ref{15}-\ref{17}, one can rewrite the above coordinates in terms of two parameters $\eta$ and $\xi$ as follows
\begin{eqnarray}
\alpha &=& -\frac{\xi}{\sin \theta},\label{29}\\
\beta &=& \sqrt{\eta - \xi^2 \cot^2 \theta + a^2 \cos^2 \theta}.\label{30}
\end{eqnarray}
Taking (\ref{23}) and (\ref{24}) into account, and choosing $\theta_{\rm o}=\frac{\pi}{2}$, we have plotted the shadow of the rotating BHs surrounded by a King-type DM halo for different values of $\rho_s$ and $r_s$ parameters in Fig.~\ref{4-shadow1}. As can be seen, at a given value of spin parameter $a$, increasing both BH parameters $\rho_s$ and $r_s$ causes an increase of the shadow of the BH. However, we note that the rate of increase in the shadow size is more pronounced in the case of fixed $\rho_s$ with varying $r_s$. Furthermore, we see that the rotating BHs in DM halo have larger shadow than the Kerr BH without DM.

\begin{figure}[H]
\centering
\includegraphics[width=3.0in]{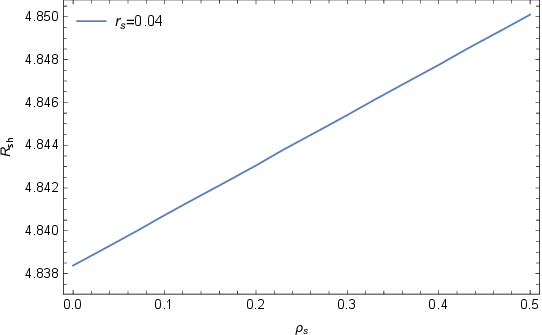}
\includegraphics[width=2.95in]{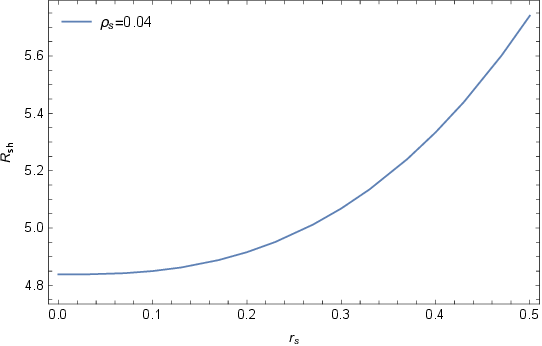}
\caption{\footnotesize The dependence of the shadow radius $R_{\rm sh}$ on the core density $\rho_s$ for $r_s=0.04$ (left panel), and on the core radius $r_s$ with $\rho_s=0.04$ (right panel). The BH spin is $a=0.9$ and the inclination angle is set to $\theta_{\rm o}=\frac{\pi}{2}$.}
\label{5-radius}
\end{figure}

The expression for the shadow radius, $R_{\rm sh}$, is given by \cite{rsh1}-\cite{rsh2}
\begin{equation}
R_{\rm sh}=\frac{1}{2}\left(\alpha(r_{\rm ph}^{+})-\alpha(r_{\rm ph}^{-})\right), \label{31}
\end{equation}
where $r_{\rm ph}^{+}$ and $r_{\rm ph}^{-}$ refer to the prograde and retrograde orbits, respectively. We have also plotted the dependence of the shadow radius $R_{\rm sh}$ on the core density $\rho_s$ and on the core radius $r_s$, in the left and right panels of Fig.~\ref{5-radius}. As can be seen, increasing both DM halo parameters $\rho_s$ and $r_s$ leads to an increase of the shadow radius, which is in agreement with the above explanation of the Fig.~\ref{4-shadow1}

\begin{figure}[H]
\centering
\includegraphics[width=2.95in]{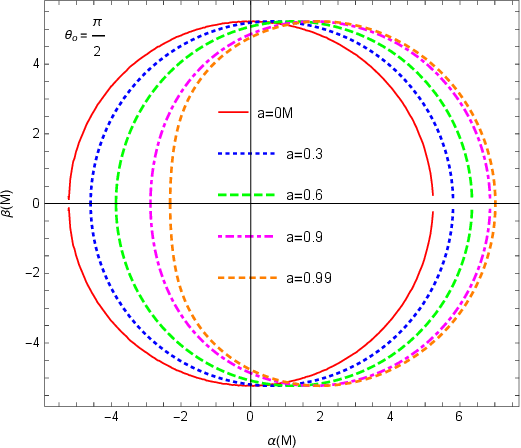}
\includegraphics[width=2.65in]{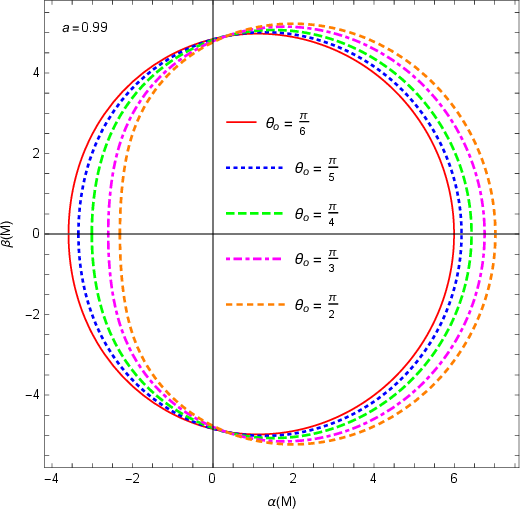}
\caption{\footnotesize The shape of shadow of the Kerr-like BH in a King DM halo for different rotation parameter with $\theta_{\rm o}=\frac{\pi}{2}$ (left panel), and for different inclinations angles with $a=0.99$ (right panel). The DM halo parameters are set to $\rho_s=0.2$ and $r_s=0.08$, with $M=1$.}
\label{6-shadow2}
\end{figure}

In Fig.~\ref{6-shadow2}, we give the shadow of the rotating BHs immersed in King DM halo for different values of the rotation parameter $a$ and for different inclination angles. The well-known effect of the BH spin on the shadow is presented in the left panel of figure, showing that, with increasing $a$, distortion of the shadow increases too. Also, from the right panel we see that, the shadow deformation increases, when the observer close to the equatorial plane.

%%%%%%%%%%%%%%%%%%%%%%%%%%%%%%%%%%%%%%%%%%%%%%%%%%%%%%%%%%%%%%%%%%%%%%%%%%%%%%%%%%%%%%%%

\begin{figure}[H]
\centering
\includegraphics[width=3.60in]{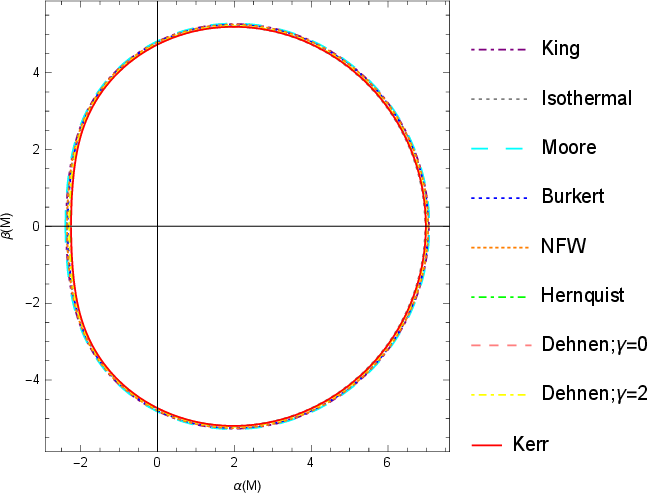}
\includegraphics[width=2.70in]{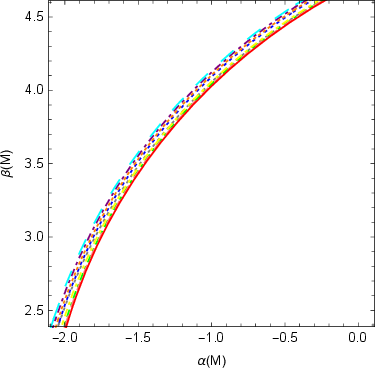}
\caption{\footnotesize The shape of shadow for rotating BHs in the King, isothermal, Moore, Burkret, NFW, Hernquist and the Dehnen DM halo models with $\rho_s=0.04$ and $r_s=0.2$ (left panel). The right panel represents zoom of the left panel, and the solid red curve corresponds to the Kerr BH in the absence of DM.}
\label{8-compare}
\end{figure}

Finally, a comparison between DM halo profiles with the Kerr BH without DM is presented in Fig.~\ref{8-compare}. As the figure shows, the presence of DM increases the shadow size; however, it has very little influence and thus the BHs immersed in DM halos are almost indistinguishable from the Kerr BH. Therefore, the effect of DM on the shadow size is negligible, and does not depend on whether the profile is cored (such as King and Burkert) or cuspy (such as Moore and NFW).  Furthermore, by examining the BH shadow one can conclude the distinction between cored and cuspy profiles is nearly impossible.

%%%%%%%%%%%%%%%%%%%%%%%%%%%%%%%%%%%%%%%%%%%%%%%%%%%%%%%%%%%%%%%%%%%%%%%%%%%%%%%%%%%%%%%%%%%%%%%%%

\begin{sidewaystable}[ht]
\centering
\caption{\footnotesize Summary of the papers presenting BHs surrounded by DM halos. Constants $\rho_s$ and $r_s$ are the characteristic density and scale parameter.}
\begin{tabular}{p{4cm} p{4cm} p{12cm} p{4cm}}
\hline\hline
\toprule
\textbf{Density distribution} & \textbf{Density profile $\rho(r)$} &\textbf{Redshift function}  &\textbf{References}\\
\midrule
\hline
Navarro-Frenk-White (NFW) \cite{Navarro1997}
    &$\displaystyle \frac{\rho_s r_s^3}{r(r + r_s)^2}$&$(1+\frac{r}{r_s})^{-\frac{8\pi \rho_s r_{s}^3}{ r}}-\frac{2M_{\rm BH}}{r}$,&\cite{Xu2018},\cite{Zeng:2025kqw} \\

Scalar Field Dark Matter (SFDM)
      &$\displaystyle \frac{\rho_s \sin(\pi r/r_s)}{(\pi r/r_s)}$&$\exp[-\frac{8G\rho_s r_{s}^2}{\pi}\frac{\sin(\pi r/r_s)}{\pi r/r_s}]-\frac{2M_{\rm BH}}{r}$,&\cite{Xu2018},\cite{Zeng:2025kqw} \\

Burkert
\cite{Burkert1995}
    & $\displaystyle \frac{\rho_s r_s^3}{(r + r_s)(r^2 + r_s^2)}$&$(1+\frac{r^2}{r_s^2})^{\frac{-2\pi \rho_s r_s^3}{r}(1-\frac{r}{r_s})}(1+\frac{r}{r_s})^{\frac{-4\pi \rho_s r_s^3}{r}(1+\frac{r}{r_s})}\exp[\frac{4\pi \rho_s r_s^3 \arctan(\frac{r}{r_s})(1+\frac{r}{r_s})}{r}]-\frac{2M_{\rm BH}}{r}$,&\cite{Zhang2021},\cite{Jusufi:2019nrn} \\

King \cite{King}
    & $\displaystyle \frac{\rho_s r_s^3}{\left({r^2+{r_s}^2}\right)^{3/2}}$
    & $\left(\frac{r}{r_s}+\sqrt{1+(\frac{r}{r_s}})^2\right)^{-\frac{8\pi \rho_s r_{s}^3}{ r}}-\frac{2M_{\rm BH}}{r}$,&\cite{Al-Badawi:2026cjx}\\

Einasto \cite{Einasto2012}
    & $\displaystyle \rho_s \exp\left[-\frac{2}{\alpha}\left(\frac{r}{r_s}\right)^{\alpha}-1\right]$
    & $\exp \left[ -\frac{\rho_s}{{\alpha}}{4\pi 2^{1-\frac{3}{\alpha}} e^{\frac{2}{\alpha}} r^2 \left( \frac{(r/r_s)^{\alpha}}{\alpha} \right)^{\frac{-3}{\alpha}}\Gamma\left[ \frac{3}{\alpha}, 0, \frac{2(\frac{r}{r_s})^{\alpha}}{\alpha}\right]}\right] - \frac{2M_{\rm BH}}{r}$,&\cite{Liu:2023oab} \\

Pseudo-isothermal \cite{Begeman}
    & $\displaystyle \frac{\rho_s r_s^2}{(r^2 + r_s^2)}$
    &$\left( r_s^2 + r^2 \right)^{4\pi\rho_s r_s^2} \exp\left[ \frac{8\pi\rho_s r_s^3 \arctan\left(\frac{r}{r_s}\right)}{r} \right] - \frac{2M_{\rm BH}}{r}$,&\cite{Yang:2023tip} \\

Hernquist \cite{Hernquist}
    & $\displaystyle \frac{\rho_s r_s^4}{r ({r}+{r_s})^3}$
    &$\left( r_s^2 + r^2 \right)^{4\pi\rho_s r_s^2} \exp\left[ \frac{8\pi\rho_s r_s^3 \arctan\left(\frac{r}{r_s}\right)}{r} \right]$
    $\displaystyle -\frac{2M_{\rm BH}}{r}$,&\cite{Jumaniyozov:2025xxh},\cite{Jha:2025xjf} \\

Dehnen-(1,4,$\gamma$)\cite{Dehnen1993}
    & $\displaystyle \frac{\rho_s r_s^{\gamma}}{r^{\gamma}(r + r_s)^{4-\gamma}}$
    &$\displaystyle 1-\frac{2M_{\rm BH}}{r}-32\pi\rho_s r_s^3\sqrt{\frac{r+r_s}{r r_s^2}}$, ($\gamma=5/2$);
    $\displaystyle 1-\frac{2M_{\rm BH}}{r}-\frac{4\pi\rho_s r_s^3 (2r+r_s)}{3(r+r_s)^2}$, ($\gamma=0$);
     $\displaystyle 1-\frac{2M_{\rm BH}}{r}-8\pi\rho_s r_s^2 \log(1+
     \frac{r}{r_s})$, ($\gamma=2$),&\cite{Al-Badawi:2024asn};\cite{UktamjonUktamov:2025emm},\cite{UktamjonUktamov:2025emm};\cite{Gohain:2025bij},\cite{Gohain:2024eer} \\

Moore \cite{Moore:1999gc}
    & $\displaystyle \frac{\rho_s r_s^3}{r^{3/2}(r ^{3/2}+ r_s^{3/2})}$
    &$\displaystyle \exp\left[{\frac{16\pi\rho_s r_s^2}{\sqrt{3}} \arctan\left(\frac{2\sqrt{{r}/{r_s}}-1}{\sqrt{3}}\right)}\right] \left(1+(\frac{r}{r_s})^{3/2}\right)^{-\frac{16\pi\rho_s r_s^3}{3r}} \left(\frac{1+\frac{r}{r_s}-\sqrt{\frac{r}{r_s}}}{1+\frac{r}{r_s}+2\sqrt{\frac{r}{r_s}}}\right)^{-\frac{8\pi\rho_s r_s^2}{3}}-\frac{2M_{\rm BH}}{r}$, &\cite{Qiao:2024ehj}\\
\bottomrule
\hline\hline
\end{tabular}
\end{sidewaystable}

%&&&&&&&&&&&&&&&&&&&&&&&&&&&&&&&&&&&&&&&&&&&&&&&&&&&&&&&&&&&&&&&&&&&&&&&&&&&&&&&&&&&&&&&&&&&&&&&&&&&&&&&&&&&&&&&&&&&&&&&&&&&&&&&&&&&&&&&&&&&&&&&&&&&&&&&&&&&&&&&&&&&&&&&&&&

\section{Conclusions}
\label{5-results}
Most of supermassive BHs reside in galactic centers, such as that of the Milky Way and M87 galaxies, whereas dwarf galaxies often do not host central galactic BHs \cite{Ghez:1998ph}. Since observations of low-surface-brightness galactic regions in dwarf galaxies tend to favor cored DM profiles, such as the King and Burkert density profiles \cite{Burkert1995}, this question arises: Can cored profiles like King be used to describe the DM surrounding a central galactic BH? Recent observations report that supermassive BHs may also exist in these dwarf galaxies; for example, the detection of a supermassive BH  in the dwarf galaxy  \cite{chandra} , or a BH with a mass  in the Leo I galaxy has been reported \cite{Bustamante-Rosell:2021ldj}-\cite{Aditya:2025lfm}. Therefore, the study of the BH solutions in galactic centers surrounded by DM halo for various DM profiles, seems unavoidable.

In recent years, many papers have presented static and rotating BHs immersed in various DM halos. In Table 1, we have summarized different DM halo profiles, the corresponding BH solutions for a galactic center immersed in DM halo, as well as references that have studied null geodesics and BH shadow in these space-times. However, for three DM profiles, King \cite{Al-Badawi:2026cjx}, Hernquist \cite{Jumaniyozov:2025xxh}, and Moore models \cite{Moore:1999gc}, the rotating BH solution and its shadow have been not investigated yet. So, in the present work, we studied shadows of theses rotating BHs and explored the effects DM halo on them, and compared the corresponding results with that of the other DM profiles, as well as with that of the Kerr BH in the absence of DM.

First, using the Newman–Janis algorithm, we have obtained rotating BHs immersed in a DM halo (King, Hernquist, and Moore), and studied the effects of the BH spin $a$, the core density $\rho_s$, and the scale radius $r_s$ on the inner and outer horizons and the ergoregion (Fig.~\ref{1-delta}-~\ref{3-ergo}). The results show that for fixed values of DM parameters, increasing the BH spin decreases the event horizon radius, while for a fixed value of spin parameter, increasing both BH parameters $\rho_s$ and $r_s$ causes an increase of the event horizon radius. Then, using the Hamilton–Jacobi equation, we studied the motion of photons near the BH and the characteristics of the BH shadow. The effects of DM halo parameters on the BH shadow have displayed in Fig.~\ref{4-shadow1} and Fig.~\ref{5-radius}, showing that for a given value of the spin parameter $a$, increasing both parameters $\rho_s$ and $r_s$ causes an increase of the shadow of the BH, so that the rotating BHs in DM halo have larger shadow than the Kerr BH without DM. We also showed that with increasing spin parameter, the distortion of the shadow gradually increases (Fig.~\ref{6-shadow2}).

Finally, we compared the BH shadow for various DM profiles and found that the presence of DM increases the shadow size in comparison to the Kerr BH without DM; however, it has very little influence (Fig.~\ref{8-compare}). Thus, BHs embedded in a DM halo are almost indistinguishable from the Kerr BH. The variation of the shadow radius of a BH surrounded by a DM halo is not necessarily unique to a specific profile; it depends on the shape of the mass distribution function and its interaction with the photon sphere radius, such that for some DM models such as PFDM \cite{Ali:2025rlb}-\cite{Shodikulov:2025xax} and Dehnen model \cite{Pantig:2022whj} with $\gamma=1$, a very slight decrease in the shadow radius relative to the Kerr BH has been reported, contrary to expectations. However, the important point is not the increase or decrease of the shadow's radius compared to the Kerr BH, but rather that the variation compared to the absence of DM is very tiny. Therefore, the effect of DM on the size and shape of the BH shadow is negligible, and does not depend on whether the profile is cored or cuspy.  Furthermore, by examining the BH shadow one can conclude the distinction between cored and cuspy profiles is nearly impossible.

%&&&&&&&&&&&&&&&&&&&&&&&&&&&&&&&&&&&&&&&&&&&&&&&&&&&&&&&&&&&&&&&&&&&&&&&&&&&&&&&&&&&&&&&&&&&&&&&&&&&&&&&&&&&&&&&&&&&&&&&&&&&&&&&&&&&&&&&&&&&&&&&&&&&&&&&&&&&&&&&&&&&&&&&&&&&&&

\subsection*{Acknowledgements}
This work is based upon research funded by Iran National Science Foundation (INSF) under project No.40404541.

%&&&&&&&&&&&&&&&&&&&&&&&&&&&&&&&&&&&&&&&&&&&&&&&&&&&&&&&&&&&&&&&&&&&&&&&&&&&&&&&&&&&&&&&&&&&&&&&&&&&&&&&&&&&&&&&&&&&&&&&&&&&&&&&&&&&&&&&&&&&&&&&&&&&&&&&&&&&&&&&&&&&


\begin{thebibliography}{99}

\bibitem{Rubin1970} V. C. Rubin and W. K. Ford,
%Rotation of the Andromeda Nebula from a Spectroscopic Survey of Emission Regions,
{\it Astrophys. J.} \textbf{159} (1970) 379.


%\cite{Corbelli:1999af}
\bibitem{Corbelli:1999af}
E.~Corbelli and P.~Salucci,
%``The Extended Rotation Curve and the Dark Matter Halo of M33,''
{\it Mon. Not. Roy. Astron. Soc.} \textbf{311} (2000) 441.
%doi:10.1046/j.1365-8711.2000.03075.x
%[arXiv:astro-ph/9909252 [astro-ph]].
%507 citations counted in INSPIRE as of 06 May 2026


%\cite{Clowe:2006eq}
\bibitem{Clowe:2006eq}
D.~Clowe, M.~Bradac, A.~H.~Gonzalez, M.~Markevitch, S.~W.~Randall, and et. al.,
%``A direct empirical proof of the existence of dark matter,''
{\it Astrophys. J. Lett.} \textbf{648} (2006) L109.
%doi:10.1086/508162
%[arXiv:astro-ph/0608407 [astro-ph]].
%3356 citations counted in INSPIRE as of 06 May 2026


%\cite{Davis:1985rj}
\bibitem{Davis:1985rj}
M.~Davis, G.~Efstathiou, C.~S.~Frenk and S.~D.~M.~White,
%``The Evolution of Large Scale Structure in a Universe Dominated by Cold Dark Matter,''
{\it Astrophys. J.} \textbf{292} (1985) 371.
%doi:10.1086/163168
%2896 citations counted in INSPIRE as of 06 May 2026


%\cite{WMAP:2010qai}
\bibitem{WMAP:2010qai}
E.~Komatsu \textit{et al.} [WMAP],
%``Seven-Year Wilkinson Microwave Anisotropy Probe (WMAP) Observations: Cosmological Interpretation,''
{\it Astrophys. J. Suppl.} \textbf{192} (2011) 18.
%doi:10.1088/0067-0049/192/2/18
%1[arXiv:1001.4538 [astro-ph.CO]].
%8311 citations counted in INSPIRE as of 06 May 2026


%\cite{Planck:2013pxb}
\bibitem{Planck:2013pxb}
P.~A.~R.~Ade \textit{et al.} [Planck],
%``Planck 2013 results. XVI. Cosmological parameters,''
{\it Astron. Astrophys.} \textbf{571} (2014) A16.
%doi:10.1051/0004-6361/201321591
%[arXiv:1303.5076 [astro-ph.CO]].
%8859 citations counted in INSPIRE as of 06 May 2026


%\cite{Massey:2010hh}
\bibitem{Massey:2010hh}
R.~Massey, T.~Kitching and J.~Richard,
%``The dark matter of gravitational lensing,''
{\it Rept. Prog. Phys.} \textbf{73} (2010) 086901.
%doi:10.1088/0034-4885/73/8/086901
%[arXiv:1001.1739 [astro-ph.CO]].
%496 citations counted in INSPIRE as of 06 May 2026


%\cite{Vegetti:2023mgp}
\bibitem{Vegetti:2023mgp}
S.~Vegetti, S.~Birrer, G.~Despali, C.~D.~Fassnacht, D.~Gilman and et.al.,
%``Strong Gravitational Lensing as a Probe of Dark Matter,''
{\it Space Sci. Rev.} \textbf{220} (2024) 58.
%doi:10.1007/s11214-024-01087-w
%[arXiv:2306.11781 [astro-ph.CO]].
%102 citations counted in INSPIRE as of 06 May 2026


\bibitem{26} J. Zavala and C. S. Frenk,
%Dark matter haloes and subhaloes,
{\it Galaxies} \textbf{7} (2019) 81.
%arXiv:1907.11775[astro-ph.CO].


\bibitem{27} J. Wang, S. Bose, C. S. Frenk, L. Gao, A. Jenkins, and et. al.,
%Universal structure of dark matter haloes over a mass range of 20 orders of magnitude,
{\it Nature} \textbf{585} (2020) 39.
%arXiv:1911.09720[astro-ph.CO].


\bibitem{28} S. Bhattacharya, S. Habib, K. Heitmann and A. Vikhlinin,
%Dark Matter Halo Profiles of Massive Clusters:
{\it Theory vs. Observations, Astrophys. J.} \textbf{766} (2013) 32.
%arXiv:1112.5479[astro-ph.CO].


%%%%%%%%%%%%%%%%%%%%%%%%%%%%%%%%%%%%%%%%%%%%%%%%


\bibitem{LIGO:2017dbh} B. P. Abbott et al. (LIGO Scientific, Virgo), {\it Phys. Rev. Lett} \textbf{116} (2016) 061102.

%\cite{Ghez:1998ph}
\bibitem{Ghez:1998ph}
A.~M.~Ghez, B.~L.~Klein, M.~Morris and E.~E.~Becklin,
%``High proper motion stars in the vicinity of Sgr A*: Evidence for a supermassive black hole at the center of our galaxy,''
{\it Astrophys. J.} \textbf{509} (1998) 678.
%doi:10.1086/306528
%[arXiv:astro-ph/9807210 [astro-ph]].
%575 citations counted in INSPIRE as of 06 May 2026


%\cite{Schodel:2002py}
\bibitem{Schodel:2002py}
R.~Schodel, T.~Ott, R.~Genzel, R.~Hofmann, M.~Lehnert, and et. al.,
%``A Star in a 15.2 year orbit around the supermassive black hole at the center of the Milky Way,''
{\it Nature} \textbf{419} (2002) 694.
%doi:10.1038/nature01121
%[arXiv:astro-ph/0210426 [astro-ph]].
%662 citations counted in INSPIRE as of 06 May 2026


%\cite{EventHorizonTelescope:2019dse}
\bibitem{EventHorizonTelescope:2019dse}
K.~Akiyama \textit{et al.} [Event Horizon Telescope],
%``First M87 Event Horizon Telescope Results. I. The Shadow of the Supermassive Black Hole,''
{\it Astrophys. J. Lett.} \textbf{875} (2019) L1.
%doi:10.3847/2041-8213/ab0ec7
%[arXiv:1906.11238 [astro-ph.GA]].
%4880 citations counted in INSPIRE as of 06 May 2026


%\cite{EventHorizonTelescope:2022wkp}
\bibitem{EventHorizonTelescope:2022wkp}
K.~Akiyama \textit{et al.} [Event Horizon Telescope],
%``First Sagittarius A* Event Horizon Telescope Results. I. The Shadow of the Supermassive Black Hole in the Center of the Milky Way,''
{\it Astrophys. J. Lett.} \textbf{930} (2022)  L12.
%doi:10.3847/2041-8213/ac6674
%[arXiv:2311.08680 [astro-ph.HE]].
%2273 citations counted in INSPIRE as of 06 May 2026


\bibitem{Synge}
J. L. Synge,
%"The Escape of Photons from Gravitationally Intense Stars,"
{\it Mon. Not. Roy. Astron. Soc.} \textbf{131} (1966) 463.


\bibitem{Luminet}
J. P. Luminet,
%"Image of a spherical black hole with thin accretion disk,"
{\it Astron. Astrophys.} \textbf{75} (1979) 228.


\bibitem{Bardeen}
J. M. Bardeen,
%"Timelike and null geodesics in the Kerr metric,"
in Black Holes (Les Astres Occlus), pp. 215-239, Gordon and Breach Science Publishers, New York, (1973).


%\cite{Kumaran:2022soh}
\bibitem{Kumaran:2022soh}
Y.~Kumaran and A.~{\"O}vg{\"u}n,
%``Deflection Angle and Shadow of the Reissner{\textendash}Nordstr{\"o}m Black Hole with Higher-Order Magnetic Correction in Einstein-Nonlinear-Maxwell Fields,''
{\it Symmetry} \textbf{14} (2022) 2054.
%doi:10.3390/sym14102054
%[arXiv:2210.00468 [gr-qc]].
%48 citations counted in INSPIRE as of 06 May 2026


%%%%%%%%%%%%%%%%%%%%%%%%%%%%%%%%%%%%
%\cite{Konoplya:2019sns}
\bibitem{Konoplya:2019sns}
R.~A.~Konoplya,
%``Shadow of a black hole surrounded by dark matter,''
{\it Phys. Lett.} B \textbf{795} (2019) 1.
%doi:10.1016/j.physletb.2019.05.043
%[arXiv:1905.00064 [gr-qc]].
%347 citations counted in INSPIRE as of 06 May 2026



%\cite{Jusufi:2020cpn}
\bibitem{Jusufi:2020cpn}
K.~Jusufi, M.~Jamil and T.~Zhu,
%``Shadows of Sgr A$^{*}$ black hole surrounded by superfluid dark matter halo,''
{\it Eur. Phys. J.} C \textbf{80} (2020) 354.
%doi:10.1140/epjc/s10052-020-7899-5
%[arXiv:2005.05299 [gr-qc]].
%124 citations counted in INSPIRE as of 06 May 2026


%\cite{Saurabh:2020zqg}
\bibitem{Saurabh:2020zqg}
K.~Saurabh and K.~Jusufi,
%``Imprints of dark matter on black hole shadows using spherical accretions,''
{\it Eur. Phys. J.} C \textbf{81} (2021) 490.
%doi:10.1140/epjc/s10052-021-09280-9
%[arXiv:2009.10599 [gr-qc]].
%118 citations counted in INSPIRE as of 06 May 2026


%\cite{Das:2021otl}
\bibitem{Das:2021otl}
A.~Das, A.~Saha and S.~Gangopadhyay,
%``Study of circular geodesics and shadow of rotating charged black hole surrounded by perfect fluid dark matter immersed in plasma,''
{\it Class. Quant. Grav.} \textbf{39} (2022) 075005.
%doi:10.1088/1361-6382/ac50ed
%[arXiv:2110.11704 [gr-qc]].
%43 citations counted in INSPIRE as of 06 May 2026


%\cite{Ma:2022jsy}
\bibitem{Ma:2022jsy}
S.~J.~Ma, T.~C.~Ma, J.~B.~Deng and X.~R.~Hu,
%``Shadow of Schwarzschild black hole in the cold dark matter halo,''
{\it Mod. Phys. Lett.} A \textbf{38} (2023) 2350104.
%doi:10.1142/S0217732323501043
%[arXiv:2206.12820 [gr-qc]].
%24 citations counted in INSPIRE as of 06 May 2026


%\cite{Hou:2018bar}
\bibitem{Hou:2018bar}
X.~Hou, Z.~Xu, M.~Zhou and J.~Wang,
%``Black hole shadow of Sgr A$^{*}$ in dark matter halo,''
{\it JCAP} \textbf{07} (2018) 015.
%doi:10.1088/1475-7516/2018/07/015
%[arXiv:1804.08110 [gr-qc]].
%125 citations counted in INSPIRE as of 06 May 2026


%\cite{Hou:2018avu}
\bibitem{Hou:2018avu}
X.~Hou, Z.~Xu and J.~Wang,
%``Rotating Black Hole Shadow in Perfect Fluid Dark Matter,''
{\it JCAP} \textbf{12} (2018) 040.
%doi:10.1088/1475-7516/2018/12/040
%[arXiv:1810.06381 [gr-qc]].
%129 citations counted in INSPIRE as of 06 May 2026


%\cite{Vagnozzi:2022moj}
\bibitem{Vagnozzi:2022moj}
S.~Vagnozzi, R.~Roy, Y.~D.~Tsai, L.~Visinelli, M.~Afrin, and et. al.,
%``Horizon-scale tests of gravity theories and fundamental physics from the Event Horizon Telescope image of Sagittarius A,''
{\it Class. Quant. Grav.} \textbf{40} (2023) 165007.
%doi:10.1088/1361-6382/acd97b
%[arXiv:2205.07787 [gr-qc]].
%899 citations counted in INSPIRE as of 06 May 2026


%\cite{Pantig:2022sjb}
\bibitem{Pantig:2022sjb}
R.~C.~Pantig and A.~{\"O}vg{\"u}n,
%``Black Hole in Quantum Wave Dark Matter,''
{\it Fortsch. Phys.} \textbf{71} (2023) 2200164.
%doi:10.1002/prop.202200164
%[arXiv:2210.00523 [gr-qc]].
%69 citations counted in INSPIRE as of 06 May 2026


%\cite{Faraji:2025efg}
\bibitem{Faraji:2025efg}
S.~Faraji and J.~L.~Rosa,
%``Shadow of quadrupole-deformed compact objects in a local dark-matter shell,''
{\it Astron. Astrophys.} \textbf{699} (2025) A266.
%doi:10.1051/0004-6361/202449922
%[arXiv:2403.02597 [astro-ph.HE]].
%10 citations counted in INSPIRE as of 06 May 2026


%\cite{Heydari-Fard:2022xhr}
\bibitem{Heydari-Fard:2022xhr}
M.~Heydari-Fard, S.~G.~Honarvar and M.~Heydari-Fard,
%``Thin accretion disc luminosity and its image around rotating black holes in perfect fluid dark matter,''
{\it Mon. Not. Roy. Astron. Soc.} \textbf{521} (2023) 708.
%doi:10.1093/mnras/stad558
%[arXiv:2210.04173 [gr-qc]].
%46 citations counted in INSPIRE as of 06 May 2026


%\cite{Bamber:2022pbs}{Karydas:2024fcn}
\bibitem{Bamber:2022pbs}
J.~Bamber, J.~C.~Aurrekoetxea, K.~Clough and P.~G.~Ferreira,
%``Black hole merger simulations in wave dark matter environments,''
{\it Phys. Rev.} D \textbf{107} (2023) 024035.
%doi:10.1103/PhysRevD.107.024035
%[arXiv:2210.09254 [gr-qc]].
%72 citations counted in INSPIRE as of 06 May 2026


%\cite{Karydas:2024fcn}
\bibitem{Karydas:2024fcn}
T.~K.~Karydas, B.~J.~Kavanagh and G.~Bertone,
%``Sharpening the dark matter signature in gravitational waveforms.~I.~Accretion and eccentricity evolution,''
{\it Phys. Rev.} D \textbf{111} (2025) 063070.
%doi:10.1103/PhysRevD.111.063070
%[arXiv:2402.13053 [gr-qc]].
%42 citations counted in INSPIRE as of 06 May 2026


%\cite{Cao:2021dcq}
\bibitem{Cao:2021dcq}
Y.~Cao, H.~Feng, W.~Hong and J.~Tao,
%``Joule{\textendash}Thomson expansion of RN-AdS black hole immersed in perfect fluid dark matter,''
{\it Commun. Theor. Phys.} \textbf{73} (2021) 095403.
%doi:10.1088/1572-9494/ac1066
%[arXiv:2101.08199 [gr-qc]].
%33 citations counted in INSPIRE as of 06 May 2026


%\cite{Zhou:2022eft}
\bibitem{Zhou:2022eft}
X.~Zhou, Y.~Xue, B.~Mu and J.~Tao,
%``Temporal and spatial chaos of RN-AdS black holes immersed in Perfect Fluid Dark Matter,''
{\it Phys. Dark. Univ.} \textbf{39} (2023) 101168.
%doi:10.1016/j.dark.2023.101168
%[arXiv:2209.03612 [gr-qc]].
%15 citations counted in INSPIRE as of 06 May 2026


%\cite{Ndongmo:2021how}
\bibitem{Ndongmo:2021how}
R.~B.~Ndongmo, S.~Mahamat, C.~B.~Tabi, T.~B.~Bouetou and T.~C.~Kofane,
%``Thermodynamics of non-linear magnetic-charged AdS black hole surrounded by quintessence, in the background of perfect fluid dark matter,''
{\it Phys. Dark. Univ.} \textbf{42} (2023) 101299.
%doi:10.1016/j.dark.2023.101299
%[arXiv:2111.05045 [gr-qc]].
%15 citations counted in INSPIRE as of 06 May 2026


%\cite{Xu:2016ylr}
\bibitem{Xu:2016ylr}
Z.~Xu, X.~Hou, J.~Wang and Y.~Liao,
%``Perfect fluid dark matter influence on thermodynamics and phase transition for a Reissner-Nordstrom-anti-de Sitter black hole,''
{\it Adv. High Energy Phys.} \textbf{2019} (2019) 2434390.
%doi:10.1155/2019/2434390
%[arXiv:1610.05454 [gr-qc]].
%48 citations counted in INSPIRE as of 06 May 2026


%\cite{Jusufi:2019ltj}
\bibitem{Jusufi:2019ltj}
K.~Jusufi,
%``Quasinormal Modes of Black Holes Surrounded by Dark Matter and Their Connection with the Shadow Radius,''
{\it Phys. Rev.} D \textbf{101} (2020) 084055.
%doi:10.1103/PhysRevD.101.084055
%[arXiv:1912.13320 [gr-qc]].
%167 citations counted in INSPIRE as of 06 May 2026


%\cite{Liu:2022ygf}
\bibitem{Liu:2022ygf}
D.~Liu, Y.~Yang, A.~{\"O}vg{\"u}n, Z.~W.~Long and Z.~Xu,
%``Gravitational ringing and superradiant instabilities of the Kerr-like black holes in a dark matter halo,''
{\it Eur. Phys. J.} C \textbf{83} (2023) 565.
%doi:10.1140/epjc/s10052-023-11739-w
%[arXiv:2204.11563 [gr-qc]].
%48 citations counted in INSPIRE as of 06 May 2026


%\cite{Liu:2024xcd}
\bibitem{Liu:2024xcd}
D.~Liu, Y.~Yang and Z.~W.~Long,
%``Probing black holes in a dark matter spike of M87 using quasinormal modes,''
{\it Eur. Phys. J.} C \textbf{84} (2024) 731.
%doi:10.1140/epjc/s10052-024-13096-8
%[arXiv:2401.09182 [gr-qc]].
%28 citations counted in INSPIRE as of 06 May 2026


%\cite{Liu:2021xfb}
\bibitem{Liu:2021xfb}
D.~Liu, Y.~Yang, S.~Wu, Y.~Xing, Z.~Xu, and et. al.,
%``Ringing of a black hole in a dark matter halo,''
{\it Phys. Rev.} D \textbf{104} (2021) 104042.
%doi:10.1103/PhysRevD.104.104042
%[arXiv:2104.04332 [gr-qc]].
%41 citations counted in INSPIRE as of 06 May 2026


\bibitem{Cardoso}
V. Cardoso, K. Destounis, F. Duque, R. P. Macedo and A. Maselli,
%``Black holes in galaxies: environmental impact on gravitational-wave generation and propagation''
{\it Phys. Rev.} D \textbf{105} (2022) L061501.



%%%%%%%%%%%%%%%%%%%%%%%%%%%%%%%
%%%%%%%%%%%%%%%%%%%%%%%%%%%%%
\bibitem{Xu2018}
Z. Xu, X. Hou, X. Gong, and J. Wang,
%\emph{Black hole space-time in dark matter halo},
{\it JCAP} \textbf{09} (2018) 038.
%arXiv:1803.00767.


%\cite{Zeng:2025kqw}
\bibitem{Zeng:2025kqw}
X.~X.~Zeng, C.~Y.~Yang, M.~I.~Aslam, R.~Saleem and S.~Aslam,
%``\emph{Kerr-like black hole surrounded by cold dark matter halo: the shadow images and EHT constraints},''
{\it JCAP} \textbf{08} (2025) 066.
%doi:10.1088/1475-7516/2025/08/066
%[arXiv:2505.07063 [gr-qc]].
%21 citations counted in INSPIRE as of 25 Apr 2026


%\cite{Jusufi:2019nrn}
\bibitem{Jusufi:2019nrn}
K.~Jusufi, M.~Jamil, P.~Salucci, T.~Zhu and S.~Haroon,
%``\emph{Black Hole Surrounded by a Dark Matter Halo in the M87 Galactic Center and its Identification with Shadow Images},''
{\it Phys. Rev.} D \textbf{100} (2019) 044012.
%doi:10.1103/PhysRevD.100.044012
%[arXiv:1905.11803 [physics.gen-ph]].
%191 citations counted in INSPIRE as of 25 Apr 2026


\bibitem{Zhang2021}
C. Zhang, T. Zhu, and A. Wang,
%\emph{Gravitational waves from the merger of a binary black hole surrounded by dark matter},
{\it Phys. Rev.} D \textbf{104} (2021) 124082.
%arXiv:2111.04966.


\bibitem{Navarro1997}
J. F. Navarro, C. S. Frenk, and S. D. M. White,
%\emph{A Universal Density Profile from Hierarchical Clustering},
{\it Astrophys. J.} \textbf{490} (1997) 493.
%arXiv:astro-ph/9611107.


\bibitem{Burkert1995}
A. Burkert,
%\emph{The Structure of Dark Matter Halos in Dwarf Galaxies},
{\it Astrophys. J. Lett.} \textbf{447} (1995) L25.
%arXiv:astro-ph/9504041.


\bibitem{Einasto2012}
E. Retana-Montenegro, E. Van Hese, G. Gentile, M. Baes, and F. Frutos-Alfaro,
%\emph{Analytical properties of Einasto dark matter haloes},
{\it Astron. Astrophys.} \textbf{540} (2012) A70.
%arXiv:1202.5242.


%\cite{Liu:2023oab}
\bibitem{Liu:2023oab}
D.~Liu, Y.~Yang, Z.~Xu and Z.~W.~Long,
%``\emph{Modeling the black holes surrounded by a dark matter halo in the galactic center of M87},''
{\it Eur. Phys. J.} C \textbf{84} (2024) 136.
%doi:10.1140/epjc/s10052-024-12492-4
%[arXiv:2307.13553 [gr-qc]].
%21 citations counted in INSPIRE as of 25 Apr 2026


\bibitem{Begeman} K. G. Begeman, A. H. Broeils, and R. H. Sanders,
{\it  Mon. Not. Roy. Astron. Soc.} \textbf{249} (1991) 523.


%\cite{Yang:2023tip}
\bibitem{Yang:2023tip}
Y.~Yang, D.~Liu, A.~{\"O}vg{\"u}n, G.~Lambiase and Z.~W.~Long,
%``\emph{Black hole surrounded by the pseudo-isothermal dark matter halo},''
{\it Eur. Phys. J.} C \textbf{84} (2024) 63.
%doi:10.1140/epjc/s10052-024-12412-6
%[arXiv:2308.05544 [gr-qc]].
%56 citations counted in INSPIRE as of 25 Apr 2026


\bibitem{King} I. King, {\it Astron. J.} {\bf 67} (1962) 471.


%\cite{Al-Badawi:2026cjx}
\bibitem{Al-Badawi:2026cjx}
A.~Al-Badawi and F.~Ahmed,
%``\emph{Spherically symmetric black hole with king dark matter halo},''
{\it Eur. Phys. J.} C \textbf{86} (2026) 139.
%doi:10.1140/epjc/s10052-026-15361-4
%2 citations counted in INSPIRE as of 25 Apr 2026


\bibitem{Hernquist} L. Hernquist,
{\it Astrophys. J.} \textbf{356} (1990) 359.


%\cite{Jumaniyozov:2025xxh}
\bibitem{Jumaniyozov:2025xxh}
S.~Jumaniyozov,
%`\emph{Thermodynamic fluctuations and radiation properties around Schwarzschild black holes immersed in Hernquist dark matter halo},''
{\it Eur. Phys. J.} C \textbf{85} (2025) 1267.
%doi:10.1140/epjc/s10052-025-15025-9
%3 citations counted in INSPIRE as of 25 Apr 2026


%\cite{Jha:2025xjf}
\bibitem{Jha:2025xjf}
S.~K.~Jha,
%``Thermodynamics, weak gravitational lensing, and parameter estimation of a Schwarzschild black hole immersed in Hernquist dark matter halo,''
{\it JCAP} \textbf{06} (2025) 033.
%doi:10.1088/1475-7516/2025/06/033
%[arXiv:2503.19938 [gr-qc]].
%17 citations counted in INSPIRE as of 26 Apr 2026


\bibitem{Dehnen1993}
W. Dehnen,
%\emph{A family of potential-density pairs for spherical galaxies and bulges},
{\it Mon. Not. Roy. Astron. Soc.} \textbf{265} (1993) 250.




%\cite{UktamjonUktamov:2025emm}
\bibitem{UktamjonUktamov:2025emm}
U.~Uktamov, S.~Shaymatov, B.~Ahmedov and C.~Yuan,
%``\emph{New analytical model of static black hole with a dark matter halo and parametric constraints through quasiperiodic oscillations},''
{\it Eur. Phys. J.} C \textbf{85} (2025) 1432.
%doi:10.1140/epjc/s10052-025-15171-0
%2 citations counted in INSPIRE as of 26 Apr 2026


%\cite{Gohain:2025bij}
\bibitem{Gohain:2025bij}
M.~M.~Gohain, D.~J.~Gogoi, K.~Bhuyan and P.~Phukon,
%``\emph{Investigating optical and ring-down gravitational wave properties of a rotating black hole in a Dehnen galactic dark matter halo},''
{\it JHEP} \textbf{51} (2026) 100539.
%doi:10.1016/j.jheap.2025.100539
%[arXiv:2508.18053 [astro-ph.CO]].
%2 citations counted in INSPIRE as of 25 Apr 2026


%\cite{Al-Badawi:2024asn}
\bibitem{Al-Badawi:2024asn}
A.~Al-Badawi, S.~Shaymatov and Y.~Sekhmani,
%``\emph{Schwarzschild black hole in galaxies surrounded by a dark matter halo},''
{\it JCAP} \textbf{02} (2025) 014.
%doi:10.1088/1475-7516/2025/02/014
%[arXiv:2411.01145 [gr-qc]].
%40 citations counted in INSPIRE as of 25 Apr 2026


%\cite{Gohain:2024eer}
\bibitem{Gohain:2024eer}
M.~M.~Gohain, P.~Phukon and K.~Bhuyan,
%``\emph{Thermodynamics and null geodesics of a Schwarzschild black hole surrounded by a Dehnen type dark matter halo},''
{\it Phys. Dark. Univ.} \textbf{46} (2024) 101683.
%doi:10.1016/j.dark.2024.101683
%[arXiv:2407.02872 [gr-qc]].
%57 citations counted in INSPIRE as of 25 Apr 2026



\bibitem{beta model} A. Cavaliere and R. Fusco-Femiano,
%\emph{X-rays from hot plasma in clusters of galaxies},
{\it Astron. Astrophys.} \textbf{49} (1976) 137.


%\cite{Wu:2024hxr}
\bibitem{Wu:2024hxr}
S.~R.~Wu, B.~Q.~Wang, Z.~W.~Long and H.~Chen,
%``\emph{Rotating black holes surrounded by a dark matter halo in the galactic center of M87 and Sgr A{\ensuremath{*}},''
{\it Phys. Dark. Univ.} \textbf{44} (2024) 101455.
%doi:10.1016/j.dark.2024.101455
%19 citations counted in INSPIRE as of 26 Apr 2026


%\cite{Liu:2023xtb}
\bibitem{Liu:2023xtb}
Y.~G.~Liu, C.~K.~Qiao and J.~Tao,
%``\emph{Gravitational lensing of spherically symmetric black holes in dark matter halos},''
{\it JCAP} \textbf{10} (2024) 075.
%doi:10.1088/1475-7516/2024/10/075
%[arXiv:2312.15760 [gr-qc]].
%18 citations counted in INSPIRE as of 26 Apr 2026


%\cite{Moore:1999gc}
\bibitem{Moore:1999gc}
B.~Moore, T.~R.~Quinn, F.~Governato, J.~Stadel and G.~Lake,
%``\emph{Cold collapse and the core catastrophe},''
{\it Mon. Not. Roy. Astron. Soc.} \textbf{310} (1999) 1147.
%doi:10.1046/j.1365-8711.1999.03039.x
%[arXiv:astro-ph/9903164 [astro-ph]].
%1198 citations counted in INSPIRE as of 26 Apr 2026


%\cite{Pantig:2022whj}
\bibitem{Pantig:2022whj}
R.~C.~Pantig and A.~{\"O}vg{\"u}n,
%``\emph{Dehnen halo effect on a black hole in an ultra-faint dwarf galaxy},''
{\it JCAP} \textbf{08} (2022) 056.
%doi:10.1088/1475-7516/2022/08/056
%[arXiv:2202.07404 [astro-ph.GA]].
%78 citations counted in INSPIRE as of 26 Apr 2026


%\cite{Qiao:2024ehj}
\bibitem{Qiao:2024ehj}
C.~K.~Qiao and P.~Su,
%``\emph{ Time delay of light in the gravitational lensing of supermassive black holes in dark matter halos},''
{\it Eur. Phys. J.} C \textbf{84} (2024) 1032.
%doi:10.1140/epjc/s10052-024-13403-3
%[arXiv:2403.05682 [gr-qc]].
%17 citations counted in INSPIRE as of 26 Apr 2026

%&&&&&&&&&&&&&&&&&&&&&& section  2


%\cite{Dekel-Zhao}
\bibitem{Dekel-Zhao}
J. Freundlich, F. Jiang, A. Dekel, and et.al.
%“The Dekel-Zhao profile: a mass-dependent dark-matter density profile with flexible inner slope and analytic potential, velocity dispersion, and lensing properties”,
{\it Mon. Not. Roy. Astron. Soc.} \textbf{499} (2020) 2912.


%%%%%%%%%%%%%%%%%%%%%%%%%%%%%%%%%% shadow radius

\bibitem{rsh1}
K.~Jusufi,
%``Connection Between the Shadow Radius and Quasinormal Modes in Rotating Spacetimes''
{\it Phys. Rev.} D \textbf{101} (2020) 124063.


\bibitem{rsh2}
Xing-Hui Feng and H. Lu,
%``On the size of rotating black holes''
{\it Eur. Phys. J.} C \textbf{80} (2020) 551.

%\cite{Vazquez:2003zm}
\bibitem{Vazquez:2003zm}
S.~E.~Vazquez and E.~P.~Esteban,
%``Strong field gravitational lensing by a Kerr black hole,''
Nuovo Cim. B \textbf{119} (2004), 489-519
doi:10.1393/ncb/i2004-10121-y
[arXiv:gr-qc/0308023 [gr-qc]].
%236 citations counted in INSPIRE as of 09 May 2026

%%%%%%%%%%%%%%%%%%%%%%%%%%%%%%%%%%%%%%%%%%%%%%%%%%%%%%%%%%%%%%%%%%%%%%%%%%%%%%%%%%%%%%%%%%%%%%%%%%%%%%%%%%%%%%%%%%%%%%%%%%%%%%%%%%%%%%%%%%%%%%%%%%%%%%%%%%%%%%%%%%%%%%%%%%%%%%%%%


%\bibitem{Jusufi} K. Jusufi, M. Jamil, P. Salucci, et al., {\it Phys. Rev.} D {\bf 100} (2019) 044012
%\bibitem{57} M. Azreg-Aïnou, {\it Phys. Rev.} D {\bf 90} (2014) 064041
%\bibitem{58} M. Azreg-Aïnou, {\it Phys. Lett.} B {\bf 730} (2014) 95
%\bibitem{59} M. Azreg-Aïnou, {\it Eur. Phys. J.} C {\bf 74} (2014) 1
%\bibitem{Walia} R. K. Walia, S. D. Maharaj and S. G. Ghosh, {\it Eur. Phys. J.} C {\bf 82} (2022) 547



\bibitem{chandra} Mini monster black hole could hold clues to giant’s growth,
%https://chandra.si.edu/press/22_releases/press_011022.html
(2024), chandra Press Room.


%\cite{Bustamante-Rosell:2021ldj}
\bibitem{Bustamante-Rosell:2021ldj}
M.~J.~Bustamante-Rosell, E.~Noyola, K.~Gebhardt, M.~H.~Fabricius, X.~Mazzalay, and et. al.,
%``Dynamical Analysis of the Dark Matter and Central Black Hole Mass in the Dwarf Spheroidal Leo I,''
{\it Astrophys. J.} \textbf{921} (2021) 107.
%doi:10.3847/1538-4357/ac0c79
%[arXiv:2111.04770 [astro-ph.GA]].
%26 citations counted in INSPIRE as of 06 May 2026




%\cite{Pascale:2025zga}
\bibitem{Pascale:2025zga}
R.~Pascale, C.~Nipoti, F.~Calura and A.~Della Croce,
%``Leo I: The classical dwarf spheroidal galaxy with the highest dark matter density,''
{\it Astron. Astrophys.} \textbf{700} (2025) A77.
%doi:10.1051/0004-6361/202555004
%[arXiv:2506.13847 [astro-ph.GA]].
%5 citations counted in INSPIRE as of 06 May 2026


%\cite{Aditya:2025lfm}
\bibitem{Aditya:2025lfm}
K.~Aditya and A.~Mangalam,
%``Can Dwarf Spheroidal Galaxies Host a Central Black Hole?,''
{\it Astrophys. J.} \textbf{997} (2026) 194.
%doi:10.3847/1538-4357/ae2d4f
%[arXiv:2512.14146 [astro-ph.GA]].
%1 citations counted in INSPIRE as of 06 May 2026


%\cite{Ali:2025rlb}
\bibitem{Ali:2025rlb}
M.~S.~Ali, A.~Negi and S.~Pant,
%``Influence of Perfect Fluid Dark Matter on Shadow Observables of Yang-Mills modified charged black holes,''
arXiv:2509.03507 [gr-qc].
%2 citations counted in INSPIRE as of 06 May 2026


%\cite{Atamurotov:2024nre}
\bibitem{Atamurotov:2024nre}
F.~Atamurotov, F.~Sarikulov, S.~G.~Ghosh and G.~Mustafa,
%``Exploring perfect fluid dark matter with EHT results of Sgr A* through rotating 4D-EGB black holes,''
{\it Phys. Dark. Univ.} \textbf{46} (2024) 101625.
%doi:10.1016/j.dark.2024.101625
%14 citations counted in INSPIRE as of 06 May 2026


%\cite{Shodikulov:2025xax}\cite{Ali:2025rlb}
\bibitem{Shodikulov:2025xax}
B.~Shodikulov, M.~Mirov, F.~Atamurotov, S.~G.~Ghosh and A.~Abdujabbarov,
%``Impact of Kalb{\textendash}Ramond fields and perfect fluid dark matter on black hole shadows and gravitational lensing,''
{\it Phys. Dark. Univ.} \textbf{50} (2025) 102096.
%doi:10.1016/j.dark.2025.102096
%10 citations counted in INSPIRE as of 06 May 2026







\end{thebibliography}
\end{document}